# Project Archetypes: A Blessing and a Curse for AI Development

*Completed Research Paper*

**Mateusz Dolata**
University of Zurich
Zurich, Switzerland
dolata@ifi.uzh.ch

**Kevin Crowston**
Syracuse University
Syracuse, NY, USA
crowston@syr.edu

**Gerhard Schwabe**
University of Zurich
Zurich, Switzerland
schwabe@ifi.uzh.ch

## Abstract

*Software projects rely on what we call project archetypes, i.e., pre-existing mental images of how projects work. They guide distribution of responsibilities, planning, or expectations. However, with the technological progress, project archetypes may become outdated, ineffective, or counterproductive by impeding more adequate approaches. Understanding archetypes of software development projects is core to leverage their potential. The development of applications using machine learning and artificial intelligence provides a context in which existing archetypes might outdate and need to be questioned, adapted, or replaced. We analyzed 36 interviews from 21 projects between IBM Watson and client companies and identified four project archetypes members initially used to understand the projects. We then derive a new project archetype, cognitive computing project, from the interviews. It can inform future development projects based on AI-development platforms. Project leaders should proactively manage project archetypes while researchers should investigate what guides initial understandings of software projects.*

## Introduction

When engaging in software development projects, especially across organizational boundaries, people seek guidance to structure the collaboration. Despite a rigid formulation of many methodologies in the literature, project participants tend to rely on best-practice solutions. As a consequence, projects are managed based on private and shared understandings of how information systems projects should work (Vlaar et al. 2008). We label frequently co-occurring understandings, models, techniques, and schemes as *project archetypes*. The application of archetypes to project management can create significant risks. First, given the role of cultural, organizational, or professional background for establishing archetypes, differences between individual mental models might lead to discrepancy and conflicts. This risk has been discussed in information systems project literature in the context of establishing mutual understanding, yet without explicit reference to archetypes as potential sources of misunderstandings (Jenkin et al. 2019; Vlaar et al. 2008). Second, since the archetypes rely on prior experience, they might become incompatible with the changing technological context and misinform the project participants about the best way forwards. At the same time, effective archetypes make the management of complex projects more successful by assuring that the involved parties are aligned in terms of goals, roles, and structure. References to archetypes might be specifically useful if a project is novel, e.g., uses new technology or moves beyond the subject expertise of the participants. The existing literature barely attends to the evolution of project archetypes based on technological changes. This manuscript explores the role of archetypes for information systems development by attending to meanings and schemata guiding the development of artificial intelligence (AI) applications.

The case of AI-based system development provides an interesting probe to study the evolution of archetypes. Whenever we refer to AI, we mean the functionality developed in the current wave of AI, i.e., features relying on inductive reasoning employed for specifiable problems and tasks (Larson 2021). State-of-the-art AI uses novel data-based, probabilistic methods involving machine learning (ML). Such functionalities are provided by AI-development platforms such as IBM Watson or Microsoft Azure AI. The platforms offer pre-trained classification and prediction models, algorithms for training models based on own data, modules for pre-processing data, and cloud computing abilities to store the data and complete computationally intensive tasks. On top, the platform providers offer support concerning the usage of the platforms and consulting services for industry partners to help them enter them build tailored AI-based applications.

AI-based applications follow a different paradigm compared to conventional deterministic software (Minsky 1991; Mitchell 2019). This difference has implications not only for how applications work but also for how they are developed. Yet, no dominating methodology for developing AI-based applications exist. On the one hand, existing methodologies and frameworks for probabilistic approaches, including CRISP-DM, are limited to the management of data-based models (Azevedo and Santos 2008) and do not embed those processes in the social or organizational context of application development. On the other hand, frameworks for application development like SCRUM or extreme programming were developed with conventional software in mind (Salo and Abrahamsson 2008; Schwaber and Beedle 2002). For instance, they assume that development of an application can be divided in meaningful chunks following a divide-and-conquer strategy. The bottom line is that there is no default method for the development of AI applications, so project partners rely on archetypes or implicit assumptions based on experience. This might impede the collective sensemaking thus jeopardizing shared understanding and the outcome of the project. To explore the archetypes and their role on AI-application development projects, we raise the following questions: (1) What are the sources of project archetypes guiding the AI-application development projects? (2) How are differences between conventional software and AI applications reflected in the project archetypes?

The notion of archetypes has been proposed to study organizational configurations and changes in those configurations (Greenwood and Hinings 1993; Kirkpatrick and Ackroyd 2003; Laughlin 1991; Schilling et al. 2017). In organizational science, an archetype is a conception "of what an organization should be doing, of how it should be doing it and how it should be judged, combined with structures and processes that serve to implement and reinforce those ideas" (Greenwood and Hinings 1988, p. 295). In this article, we adapt this notion to information systems (IS) projects, which we understand as a form of organization. Like organizations, projects can be characterized in various dimensions and values associated with the dimensions.

We apply the archetype lens to analyze 21 projects between IBM Watson and client companies. Initially, the informants relied on four project archetypes: agile software development, customization and integration, design thinking, and big-data analytics. However, these archetypes were invalidated in the course of the project and needed to be updated. The learnings participants made throughout the projects point to a new, emerging archetype for development of AI-based applications based on AI development platforms. We call the new archetype a *cognitive computing project* and compare it systematically to the previous archetypes.

From this research, practitioners learn about struggles and learnings reported in the analyzed data and obtain new, unique characteristics of AI application development projects. Also, we indicate that project methodologies and techniques are not ambivalent towards technology. This questions the claims about the potentially unlimited applicability of such paradigms as agile software development or design thinking. Researchers benefit from the proposed notion of a *project archetype* that might be applied to analyze various project and collaboration contexts. Sensemaking researchers learn where meanings initially come from and how new meanings are established based on ongoing, collective experience. Finally, AI researchers learn how differences between deterministic and probabilistic software are framed by developers.

# Related Work

## *Development of Deterministic and Probabilistic Applications*

What is currently called state-of-the-art AI uses inductive reasoning based on large data sets (Larson 2021; Mitchell 2019). This reasoning is frequently opaque, such that human users or developers cannot easily comprehend what the machine does and how it arrives at its conclusions (Molnar 2020). This way of reasoning has been specified as analogical, connectionist, and scruffy as opposite to the logical, symbolic, and neat character of deterministic software (Minsky 1991). The functionality of a deterministic system can be described as a sum of functionalities of its components. The functionality of an AI application instead emerges from a interaction between its components such as data, preprocessing modules, models, and learning algorithms, as well as inputs from the user, which can be used to retrain the model. AI applications can yield different output at different points in time. This indeterminacy makes AI applications more complex than deterministic ones and requires a robust development methodology and project guidance.

Traditional software development, yielding deterministic tools for businesses and users, offers a range of process models, such as waterfall, incremental development, the spiral model, or various agile approaches (Cockburn 2002; Dittrich 2014; Larman and Basili 2003). Similarly, a product-development perspective offers tools and guidance, with approaches such as prototyping (low-to-high-maturity), foresight analysis, or various flavors of design thinking (Dolata and Schwabe 2016; Dorst 2011; Häger et al. 2015; Uebernickel et al. 2015). Some of the listed approaches can deal with uncertainty in specific areas. For instance, agile methods like SCRUM or extreme programming can deal with uncertainty regarding technology and resources (Schwaber and Beedle 2002), while design thinking tries to manage uncertainty of requirements and user needs (Häger et al. 2015; Uebernickel et al. 2015). However, none of those models explicitly addresses the uncertainty of data and insights that results from the processing of big data. AI applications development results in probabilistic software, i.e., one that computes the probability of some outcome rather than a certain answer. The typical challenges result from the vast and continuously growing amount of mixed-quality data and the non-deterministic results generated by the application. Existing development models for software development fall short of managing those uncertainties (Konar 1999).

The same problem holds for more general project management approaches, like the Project Management Institute (PMI) standards codified in PMBOK (PMI 2017). Those standards approach the management of complex systems development using platforms, involving platform vendors as partners, and serving networks of stakeholders. However, the link between data analytics and development is missing (PMI 2017). Furthermore, models from PMI are continuously challenged by practice: companies apply software development methods and guides in a large variety of ways and deviate from them to reflect organizational structures, project context, intended products, possessed skills and abilities, etc. (Salo and Abrahamsson 2008). Even though IS research has dealt with platform-based and client-vendor development projects, the results remain explanatory rather than prescriptive and do not explicitly attend to characteristics of AI (Bonina et al. 2021; Foerderer et al. 2019; Vlaar et al. 2008; Williams 2011). Yet, a consistent guidance would be of use for AI application development given that it heavily relies on use of AI development platforms like IBM Watson or Microsoft Azure AI and involves cross-organizational cooperation and consulting services.

Finally, there are methodologies proposed for use in data analytics and data mining such as CRISP-DM (Azevedo and Santos 2008). Data analytics copes with integration of data from various sources, drawing inferences, and making predictions (Brynjolfsson et al. 2014; Gudivada 2017; Pearl and Mackenzie 2018). Today's methods for data analytics evolved from simple SQL analysis (Gudivada 2017). Simultaneously, the functional sophistication and scalability grew (Gudivada 2017). Whereas prior to big data, analytics dealt primarily with structured, relational data model, big data brought about a variety of data types, formats, and sources of large volumes (Gudivada 2017; Larson 2021). A typical methodology for a big data project involved steps like identification of a desired insight or a set of insights with relevance to business, understanding and preparation of the data, modelling and evaluation, and deployment (Azevedo and Santos 2008; Gao et al. 2015). Some processes consider integration with the existing systems or cross-functional team formation for the data examination, but the ultimate goal is a data-based insight (Dutta and Bose 2015; Kandel et al. 2012). As a consequence, methodologies and guidance for big data projects deal primarily with data issues, like data quality, amount, or data access and infrastructure (Reamy 2016).

The focus of data analytics methodologies is different to what many companies want. Rather than an insight or array of insights, they aim at applications which use insights from one or several data sets to generate organizational value and can be integrated in business processes (McAfee and Brynjolfsson 2017). Whereas an insight might be of great relevance for making decisions (e.g., measure of a key performance indicator like brand popularity based on online data), AI applications go beyond this and include analytical and generative aspects (e.g., an application providing assessment of client's trustworthiness in an underwriting process). Data-analytics methodologies fail to provide holistic guidance for building such applications that explicate project roles and responsibilities, instructions for managing unexpected difficulties, or solutions for uncertainties beyond the ones affecting the data. Practices of software engineering and development in big-data projects remain largely unattended (Madhavji et al. 2015; Otero and Peter 2015; Reamy 2016). More recent literature acknowledges the role of data scientists in software development (Kim et al. 2016; Muller et al. 2019), but their implications concern the integration of new type of expert and their skills in the projects rather than the overall guidance affecting all project members.

In summary, AI-based systems differ from deterministic systems concerning their complexity, comprehensibility, and, thus, manageability. These differences come along with various types of uncertainties concerning technology and resources, requirements and user needs, business value and impact, data quality and accessibility, lack of control and understanding of the results, as well es dependency on a platform provider. Standard, wide-spread methodologies for managing software projects fall short of accommodating this range of uncertainties. There is no "typical" or "default" way of launching and running an AI-application development project. Yet, to engage in complex environments humans must be able to make sense of what happens around them and act according to the meanings they developed (Weick 1988, 1995). In the absence of other cues, people tend initially to rely on pre-existing organizational schemas, social defaults, or collected experiences (Weick 1988). However, given the lack of dominating methodology that could provide widely accepted schemas or defaults and given that only a limited number of people from non-IT businesses have experience in AI application development, we expect that those involved in AI-system development will have to draw on other meanings to guide their behavior. We seek to explore those meanings (1) to learn about their effectiveness for the management of the AI application development projects and (2) to better understand the initial phase of sensemaking processes in individuals and collectives confronted with new technology in project context. We use the notion of archetypes to frame the study and the findings.

### *Archetypes*

The notion of an archetype emerged within organization science to describe and systematically study organizational configurations and their dynamics (Greenwood and Hinings 1988). The configurational perspective assumes that an organization as a whole is best understood as a constellations of interconnected elements (Fiss et al. 2013). Archetypes provide a way to describe common patterns of those constellations through considerations of significant dimensions of those constellations and rationale behind them (Laughlin 1991; Sawyer 2004). More specifically, an archetype characterizes the *structural arrangements* embodied in the management systems and practices, as well as organizational structures (Greenwood and Hinings 1993). Additionally, an archetype comprises *interpretative schemes*. Those include the values, beliefs, and ideas that lead to establishing the structural arrangements, i.e., stakeholders' perception of "what [an organization] should be doing, how it should be doing, and how it should be judged" (Greenwood and Hinings 1988, p. 295). Consequently, an archetype encapsulates a holistic view of an organization pattern.

he notion of an archetype was originally proposed to uncover and classify patterns of organizing and IS has used this notion in accordance with the original intention. Researchers adapted the concept to explore business models in logistics (Möller et al. 2019) or data analytics (Hunke et al. 2020; Kayser et al. 2021). For instance, they identified new types of services which emerged around outsourcing of data analytics (Hunke et al. 2020). This indicates the value of archetypes as a framing for differentiating between old and new forms of organizations. Additionally, the notion of archetypes was leveraged to popularize architectural perspective on organizations (Haki 2021; Schilling 2018; Winter 2016). In such research, archetypes describe whole (large) organizations to establish links between organizational objectives and structures. The notion of archetype is increasingly used in IS to classify and describe organizational phenomena.

Outside of IS, the concept has been mostly used to study conditions under which organizations move from one archetype to another (Greenwood and Hinings 1993; Schilling et al. 2017). Such moves comprise not only restructuring the organization and establishing new practices, but also shifts concerning the values and beliefs about the objectives and ways to achieve those objectives or values. Therein, the research focused mostly on identifying triggers for change described as environmental or contextual pressures such as globalization, (de-)regulation/change in government policy, change in client needs, technological progress, and capacity for action (Schilling et al. 2017). Yet, little attention was given to the archetypes themselves and the social processes involved in shifts (Kirkpatrick and Ackroyd 2003; Schilling et al. 2017). Specifically, it remains unclear how individual interpretative schemas emerge and impact the process of organizing.

In this article, we propose the notion of *project archetypes*. We see a project as a specific form of organization oriented towards a specific set of goals. A project is a temporary organization, one that terminates after achieving the goals or at a specific point in time. It might also come to an end upon agreement of the involved parties. In the organizational setting, especially when two or more organizations are involved, projects are a collaborative enterprise. They involve project members who might be assigned roles and responsibilities based, e.g., on their organizational affiliation or specific skills and knowledge. Aligned with the original idea of an archetype from the organizational discourse (Greenwood and Hinings 1988, 1993; Laughlin 1991), a *project archetype* attends to structural and interpretative aspects. A project archetype characterizes the structure of a project embodied in the management structure, collaboration practices, distribution of roles, tasks, and responsibilities, and rules related to participation and membership in the project. Simultaneously, a project archetype provides an interpretative schema which explains values, beliefs, and ideas which drive the structural construction. Interpretative schemas include information on the type of outcomes to be expected from the project, adequate ways to produce the outcomes, and basis for judgements concerning the project. Given the collaborative character of projects, it is essential to differentiate between individual goals of the members and a shared goal which unifies the efforts (Briggs et al. 2006). A project archetype is therefore a holistic description of a project's configuration and workings.

We use the notion of project archetypes in combination with the sensemaking perspective (Helms Mills et al. 2010; Weick 1988, 1995; Weick and Sutcliffe 2015). Sensemaking has emerged as a lens to study socio-cognitive processes and human action in *complex* situations (Holt and Cornelissen 2014; Jensen et al. 2009; Weick 1995; Weick et al. 2005). Sensemaking is the ongoing, more-or-less conscious production of plausible images or stories about what is happening to inform the subject's own action. It happens through attributing meaning to a particular target: an object, a situation, or a phenomenon that provides salient cues. Only through the perception and interpretation of those cues, captured as tentative meanings, are humans able to interact with the environment (Pratt 2000). Engagement with the environment starts with *expectations* based on past experiences (Stigliani and Ravasi 2012), inspirations, frames (Jensen et al. 2009), or preconceptions (Weick 1988). According to the sensemaking perspective, in collaborative situations the multiple players form their own initial understanding of what is happening and what should be happening. Through interaction with the environment including the social environment, they adjust their initial meanings. In the best case, the collaborators will arrive at a set of shared meanings formed through interaction with other individuals and the environment (Crowston and Kammerer 1998).

We claim that projects are typically seen first through the perspective of archetypes. Projects are frequent, recurrent, and basic form of enterprise in many organizational environments. Therefore, it is natural that archetypes emerge that provide a holistic model of a project encapsulated in a simple category. The application of archetypes can be observed in language, as actors refer to categories such as research projects, development projects, manufacturing projects, etc. The existence of archetypes can have major impact on individual and collective sensemaking. When confronted with a new phenomenon, people rely on existing frameworks "such as institutional constraints, organizational premises, plans, expectations, acceptable justifications, and traditions inherited from predecessor" (Weick et al. 2005, p. 410). Project archetypes provide such useful frameworks for an individual. Given that many project archetypes are broadly known, one can assume that other project members will know and possibly follow them too. However, when the sensemaking process remains implicit, project members might arrive at different frameworks and thus try to implement incompatible visions of a project. Development of AI-based applications is a new, still emerging phenomenon. Many individuals are engaging in this type of enterprise for the first time. It is thus necessary to understand what archetypes influence their initial perspectives and how those perspectives shift based on how the projects develop. Examination of these shifts can uncover characteristics of sensemaking of collaborative projects and so inform future AI-based development.

## Method

*Study setting*. Our study is set in the context of IBM Watson project development in Switzerland. Watson is a development platform including business-ready AI tools and solutions designed for use in development of AI-based business applications. It emerged by modularization, re-training, and extension of a question-answering engine known for its successful participation in a TV quiz show in 2011 (Mitchell 2019). Shortly thereafter, IBM started projects with other client companies to leverage the abilities in work-related contexts. In 2013, IBM opened the DP for use by independent developers and since then has continuously extended its functionalities, added new APIs, tools, and models.

In parallel, IBM engaged in commercial collaborations with organizations from around the world to identify and develop use cases for the application of Watson. Those projects address the specific needs of the client and rely on the analysis and use of the client's data sets. They include consulting as well as development services. The client pays for the services offered by IBM though hour rates and other agreements are not open to the public. Official statistics about Watson projects are not public, but IBM claims on their website that over 100 million users benefit from Watson. In these projects, IBM takes the role of a vendor. Accordingly, it provides knowledge and resources to support the client in developing an application based on the development platform it provides. As we started collecting data in 2017, IBM already had much experience with Watson projects around the world. According to internal information, IBM Switzerland had about 3 years of experience in running Watson projects and over 50 projects running or recently completed.

*Study design.* This paper follows a qualitative research methodology. We strive to understand what considerations direct people participating in those project, i.e., what are their understandings of how the projects should be configured (Yin 2003). We rely on data, observer, and theory triangulation to enhance the precision and accommodate for a broader picture of the studied phenomenon (Runeson and Höst 2009).

*Data elicitation.* Data for the study comes from interviews with informants from IBM and from its partners, to collect different opinions on the cases. To select interviewees for this study, two senior IBM managers scanned all IBM Watson projects in Switzerland, resulting in 21 selected projects involving 17 industry clients. The projects between IBM and clients combined three goals: yielding an AI-based application for use by the client, investigating potentials of long-term business cooperation, and giving the client hands-on experience with AI and Watson. For instance, a major Swiss insurance company envisioned an application that would help its underwriting department collect and summarize their own and publicly available data on small businesses to predict their risk levels and provide a more adequate insurance offering.

Between March and May 2017 our team carried out 36 semi-structured interviews with members of the selected projects. The interviewees were IBM-side project managers, client-side project managers, developers, or IBM consultants. In 17 cases, we conducted interviews with representatives of the client and IBM. Since one company was involved in two different parallel Watson projects and another company was involved in three parallel projects, interviewees from those companies reported on all projects in their interviews. Client representatives were not available in the remaining cases, so we only interviewed the IBM side. Three client-side interviewees were women; eleven were men. Five IBM-side interviewees were women, 17 were men. Employees from all organizations reported that they had previous experiences in client-vendor collaborations. All interviewees had at least two years of experience working either for IBM or the client companies, so they knew the context of their work.

To guide the interviews, four main areas of interest and multiple open questions were prepared but dynamically re-arranged depending on the conversation (Runeson and Höst 2009). The four areas were application domain, project management, requirements for AI-based development, and impact on individual/human-computer interaction. They reflected our intention to understand the projects as a whole and identify pressures that might pose challenges in those projects. Yet, they were broad enough to provide material for exploring unexpected relationships. This approach allowed for improvisation and deeper insight.

All interviews lasted at least one hour, with persons involved in more than one project, proportionally longer. Seven of the interviews were conducted in English and 29 in German. All interviews were audio-recorded, transcribed (intelligent verbatim – the transcription represents recorded speech but without fillers and repetitions that may distract the content), and offered to the subjects for review. To improve observer triangulation, we had two interviewers/coders supervised by three experienced researchers and two higher management persons from IBM to improve observer triangulation. Observations were discussed in multiple meetings throughout the data collection and analysis to support triangulation in the research team.

*Data analysis.* Data were coded in two rounds. The initial coding round was conducted bottom-up within the mentioned areas of interest and yielded approx. 3000 relevant segments. The results of the initial round were summarized and discussed in two workshops involving the researchers and IBM managers in 2017 and 2018, and two further workshops among researchers in 2018. The analysis of the initially coded segments revealed that the project members' initial understanding of what should be done in the projects, how it should be done, and what are the measures of performance changed over time.

The researchers observed similar effects in their own AI-based project work with external industry partners. They observed that the AI-based projects experienced significant tensions concerning expectations of the involved parties, which required significant effort to be resolved. Additionally, the researchers discovered that industry partners lacked understanding of how to run AI-based development projects and that exchange between industry players about insights for managing such projects was lacking. Inspired by those observations, we decided to revisit previously collected data in a systematic way to identify a potential reason for the observed tensions. This perspective informed the second round of coding.

The second round of coding considered whole interviews again but followed a top-down process. We focused on identifying and analyzing passages presenting participants' initial expectations concerning the project configuration and the insights they collected throughout the project. Those passages frequently had the form of self-reflection on the assumptions interviewees made at the beginning and how those assumptions were proven wrong, incomplete, or inadequate compared to the emerging configuration of the project. The second-round coding yielded 120 coded segments, which form the basis for the current manuscript. Two coders were involved in data analysis; they employed iterative coding. A third researcher controlled and corrected the coding, to establish a coherent basis for analysis. As all coders were bilingual, they coded the transcripts in the original language. Selected quotations have been translated for presentation.

## Findings

The material offers insights on the nature of AI-based development as seen by the project members. It points to issues typical for industry projects such as: need for a sponsor, unpredictable team dynamics, technology issues, coordination issues, and the complex nature of failure and success. However, the insights go beyond that: the interviewees vary strongly in how they initially framed the projects: they find similarities with known models or approaches, but then also identify essential differences they learned throughout the process. This section attends to the interpretative schemas participants used to make sense of the projects in the initial phase and then indicates how those schemas were invalidated and replaced later. Table 1 summarizes the collected statements and indicates the four major archetypes driving participants behaviors in the early phases of the projects. It also includes the description of the new archetype, *cognitive computing*, called after the phrasing used by IBM to market Watson platform. The dimensions we use to characterize the archetypes emerged in a bottom-up process, based on the interviews. They address aims, artefacts, activities, abilities, and assumptions. The last row lists the advantages and disadvantages of each archetype in the process of making sense of the analyzed projects, i.e., how the archetype helped framing and understanding the project vs. how it mismatched the reality demanding additional sensemaking effort.

### *Agile Software Development*

The development of Watson-based applications happens between IBM and a corporate client. Some interviewees frame this collaboration as **agile software development**. An IBM consultant clarifies: *"We have applied agile mode to this project. This means that we have been working in sprint mode. From the experience, we always or almost always take two-week sprints and at that time we not only had reviews with the client, but also daily standups or daily calls with the client. And we also had SCRUM boards with the tasks for us and for the partner people who were involved as well. We handled everything transparently"* (3V1). Other project members who recognize agile development in Watson projects emphasize their quick character and tangible output. The client-side project manager from the same project says: *"I liked the approach of how we did it in an agile approach. We delivered something very concrete, very quickly"* (3C1).

The participants describe typical activities and artefacts involved in agile software development and indicate their importance for AI-based projects. An IBM consultant lists several such aspects: *"In the agile world, you need certain preparations anyway, in the sense that you already know approximately what kind of solution you need in the end. If I now take the architecture, for example, we are talking about the system context, what kind of surrounding systems do we have, components, interfaces, software, hardware; that has not changed. But what has changed a lot in an agile approach (and we would like to do all our Watson engagements also after agile implementation) that the customer has a possibility to tell us if we are going in a right direction or not. Because in Watson it is so that very often the customer already has certain ideas or expectations at the beginning, what will or can come at the end"* (3V1). A client representative states what expertise is necessary to run agile projects: *"You need system engineer, package infrastructure specialist, UX developer, project management..."* (12C1). The collected statements add up to a notion of agile development which embraces iterative process, frequent interaction with the client, and quick and frequent output which can be tested against client's expectations.

|  | Agile Software Development | Integration, Customization, Implementation | Design Thinking Project | Big Data Analytics | Emerging Archetype: Cognitive Computing |
|---|---|---|---|---|---|
| **Aim** | Developing a unique tool for the client company | Adapting an existing platform for the client company | Innovating at the intersection of feasibility, viability, and desirability | Analysis of client's data; interpreting the data to gain insight | Research and development of a probabilistic tool in collaboration with the client company |
| **Initial Step** | Specifying requirements and developing a proof of concept | Choosing an existing platform and identifying affected processes | Identifying pain points of the company and its members | Overview of the available data and identifying knowledge targets | Identifying potentials of the available data in combination with the development platform |
| **Central Artefacts** | Prototypes, Use Cases, Sprints, Software, Architecture, Programming Infrastructure, Components, Hardware | Specifications, Platform, Modules, Business Process, Change Requests, User Exits, Reference Frameworks | Low Fidelity Prototypes, Personas, Empathy Maps, User Interfaces, Ideas, Wire-Frames, Benchmarks | Structured Data, Patterns, Statistics, Algorithms, Insight, Features, Key Indicators, Correlations | Structured and Unstructured Data, Benefit Case (benefits client expects from the new application), Platform, ML Models, Application, |
| **Core Activities** | Defining use cases, iterating, improving the prototype, sprints, daily standups, client calls | Specifying the vision, identifying adequate modules, customization based on client data | Iterative prototyping, ideation, brainstorming, workshops, interviews, evaluating, observing | Data aggregation, data analytics, data mining, insight generation, insight formulation | Improving data, managing data, (iterative) training and testing of models, generating data, learning, integrating |
| **Core Assumptions** | Client has identified a specific, unique problem or improvement potential in their company and requires a dedicated solution to address it | Product/Platform exists, it is known that it can address client's needs, and what is the effort of adapting it to the client's business processes and technical infrastructure | Clients have problems (known and yet unknown) which should be solved; it's about specifying the problem and exploring solution space to create or select the solution | Clients have access to data which potentially includes valuable insights to improve the working of the company; it's necessary to extract those insights from data | Clients have problems or needs for which a probabilistic application can provide a solution; it's about exploring ways of creating re-usable end-user applications which use the available data to address those problems |
| **Typical Roles / Abilities** | System engineer, package infrastructure specialist, UX developer, project manager, business analyst, process engineer | Business consultant, technology consultant, process analyst, IT department, developer, product owner, solution provider | Facilitator, user, client, visionary, technology expert, domain expert, developer, designer, test person | Data scientist, data analyst, data owner, database engineer, business consultant, consumer researcher | Data Consultant (competent in data processing technologies and business communication, analyzes dependency between work processes and potential of the available data) |
| **Abilities Distribution** | Client knows the problem and potential for improvement; Client knows about its own business; Vendor knows about available technologies and has experience in developing software | Client knows their vision and desires; Client knows own infrastructure for integration; Vendor / Platform provider knows about the abilities of the specific technology and has experience in adopting it; | Vendor knows about processes to find a solution and about potential technologies; Client knows its business; Users experience their problems but need help to explicate them | Vendor knows technologies for data analytics; Client has the data, knows the semantics of the data, assumes that the data includes insights and can interpret the insight in terms of business value; | Vendor knowns about general abilities of the platform but not about abilities in the connection with the given data; Vendor knows what data limitations might emerge; Client has data but knows it in a 'human' way only, without their value for ML |
| **(Dis-)Advantages for guiding AI-application projects** | **PRO**: identifies typical roles and processes; helps managing dynamic expectations; **CON**: trust in abilities of the provider depends on success or failure of ML training efforts, which are hard to recapitulate; requires divide-and-conquer. | **PRO**: accounts for expenses and roles related to selecting, customizing, and integrating external solution; **CON**: incompatible with the need for frequent contact with business users; obfuscates amount of work related to providing a module. | **PRO**: moves focus to business problems; provides an effective toolset for creative work episodes; **CON**: creates contradictions between the open-ended view of design thinking and commitment to a specific technology/data set. | **PRO**: focuses on the available data and statistical techniques as most likely sources of unexpected outcomes; **CON**: incompatible with the objective of the projects to establish a working application rather than producing an insight from the data. | **PRO**: combines data-oriented with product-oriented approaches; identifies necessary roles and capabilities (e.g., data consultant) yet allows for collaborative enacting of the roles across organizational boundaries; accounts for the importance and the open-ended character of data-based training activities; |

**Table 1. Project archetypes identified in Watson-based application development**

Yet, many interviewees who initially embraced this archetype indicate its limitations for Watson projects. For instance, the planning capability in Watson projects is lower than in other agile projects. A client representative explains: *"[In Watson projects] it is simply common for unforeseen expenses to arise again and again, problems that need to be solved, things that need to be improved and stabilized"* (11C1). A client-side project manager specifies it further: *"That's a bit of a problem, this traceability, why did they need ten days now. That is still difficult, you must trust them. You're not used to that as a large company. Normally you have a clean accounting, 'so and so much for this and that' and can understand their working hours. With the Watson, there is also a lot of training, so develop, train, try out, 'uh does not work yet', another data set, train again… And if you have a bug [in a non-AI application], you can find it and debug it. And it is easier to understand where the money goes to"* (5C1).

An IBM project manager emphasizes the role of time and the incremental development in AI-based development. He compares the AI-based development to a journey without clear boundaries: *"We now know we need to approach the topic step by step, we need to build know-how, we need to get the partner started and it's not a normal IT project with a release point – it's a journey. (…) That's one of the key learnings from this project. I've said that earlier, it's not just a big bang and then it's going, it's really a way. And, I think, so you can also build trust by simple cases, small cases that have only a minimal benefit, but that are already a step in this direction"* (2V1). This comment calls for quick release cycles and agile mindset but identifies users' skepticism and misunderstanding as an issue. Overall, the interviewees explicate the awareness, that when developing AI-based applications generates new issues, where classic tools of agile development (e.g., system evaluation against development goals and planned releases) come too short.

### *Integration, Customization, Implementation*

Some of the interviewees compare Watson projects to **software customization and deployment** using reference to other enterprise solutions. They see the platform as a portfolio of off-the-shelf applications that just need to be adjusted to work with client's data. An IBM consultant confirms this view. He focuses on specific tasks within Watson projects and sees similarities to software deployment: *"But, basically, the project approach does not change compared to other projects. In the sense that is like with any word processors [you adapt and deploy]. You need one that knows the dependencies. Then you know what kind of people you have in the project and who does what and until when. That does not change…"* (3V1).

Following this line, a client representative describes preparations for the project and how those preparations were driven by the integration and customization archetype: *"Usually, the requirements are specified together with the [internal] customer, and they were already available in this project. This was a tender and the requirements were clearly defined as to what was needed. That was also prioritized. Many things were delivered out of the box and some things had to be developed. That was the interaction between us and IBM, who had to develop something. Once it is clear what needs to be done, from a project management point of view, you certainly have to buy the services, but you also have to take care of the software licenses and then set up the system, which we then accompanied. I have to say, as Project Manager, I actually had to intervene very little. It's actually a simple thing. Depending on where Watson is implemented in the company, there is a corresponding coordination effort with the operator. Where are the servers located, network issues to be resolved, etc.?"* (21V1).

Overall, Watson projects may resemble buying software and integrating it into company's infrastructure. Yet, IBM representatives frequently assess this interpretation as insufficient for embracing the complexity of Watson projects. An IBM consultant with experience in several Watson projects summarizes his own reflection as follows: *"I mean, now that I think about it, they really didn't realize how much work it is. No customer is aware of how much work a Watson project is; all this cognitive AI stuff, everything that falls under this innovative IT [requires much effort]. Most people, most clients don't realize how much work is really behind it. They always think it's just like in the old days, when you put a mainframe there, you plugged it in. Then a small IBM person came, did a bit of tinkering, and then went home again. And here [in Watson projects] you have a lot of contact with the client. So, every day there's something you must discuss, something you have to look at, something you have to change. And that wasn't the case with the classic IT implementation projects"* (7V1). This informant contrasts perceptions concerning the necessary effort, adequate process, and evaluation criteria between Watson project and traditional integration project.

Additionally, lack of experience causes difficulties in assessing the effort related to Watson projects. An IBM-side project member concludes: *"That means, if you compare it with, for example, an SAP implementation, which has already been done a few times over the last decades, it is easier to estimate what it takes, how much it takes to achieve such a goal, and even the procedure that is used. Here you have to deal with cognitive computing with a couple of things. On the one hand, the technologies are new, and their use is new, which means that there is a lack of empirical values for the most part, and then there are topics such as machine learning and training, which has a lot to do with the quality of the data and also with the dimensionality of the data. This also makes the training not very predictable, when you will get results that are acceptable for a customer"* (8V1). This comment claims that inherent characteristics of the technology invalidate the archetype of integration, customization, and implementation. Overall, some interviewees perceive the projects through the perspective of earlier projects related to enterprise resource planning systems or office software. They think of including company's data in terms of adaptation or customization of Watson modules. Yet, this does not align with the role of data for establishing the AI functionality.

*Project Archetypes for AI Development*## Design Thinking Project

Many interviewees compare Watson projects to innovation projects implementing the paradigm of **design thinking**. An IBM-side manager argues for design thinking as the adequate tool for identifying use cases and sees it as IBM's competence: *"That's a design thinking approach. It's not that we take any use cases from Watson's brochure and present them to the customer. This does not work. Actually, you have to tell the customer: 'What are your problems that you need to solve?'. Then there are those where Watson helps and where Watson is the wrong technology. We then focused on the issues, where Watson offers a solution"* (17V1). Another IBM consultant emphasizes the role of design thinking as a toolset for the whole project - for various phases and for different tasks within those phases, including prototyping: *"Exactly, we have a design thinking method that, I think, is very well based on the theory. It starts with Ideation Workshops, Persona, Empathy Map, Map Scenario, As-Is, To-Be and, as fast as possible, prototype developments. In another project of mine, it was also very strong prototype-based"* (3V2).

The client-side interviewees confirm that design thinking drives the Watson projects. A manager from a large insurance provider emphasizes the role of design thinking workshops for problem specification. Additionally, she points out that IBM's design thinking competence is not unique anymore: *"So, we also did some kind of design thinking workshops, so brainstorming (…) and then really asking, 'what are the pains?' and 'how are they precisely?', 'who is ultimately the user of this application?' So, customer and user perspective. That helped us most. The [insurance company] is already on this topic for a long time anyway. As a company we possess a castle, a beautiful castle in [town name]. There we work with Swiss SMEs on innovation, whether they are our customers or not. So that's the Californian design thinking method we've just adapted for Switzerland. We always do workshops. Interestingly, we have also done the same with IBM. They wanted to sell us something big, but I said, we know it, it's an old hat"* (12C1).

All in all, interviewees who embrace the innovation-projects archetype attribute the design thinking character to the usage of creativity and design methods like brainstorming or personas, as well as to the focus on problem identification as opposite to solution implementation. However, there are reports of episodes or attitudes which are atypical for the open-ended design thinking paradigm. The interviewees report that design thinking workshops or activities are conducted to find the adequate problem for the provided solution as opposite to finding real pain problems and then looking for potential paths to address them. An IBM consultant explicates the process of identifying use cases to launch a Watson project – he points to feasibility as important choice criterion: *"We have done design thinking workshops. And, there, we identified these use cases together with the customer. Back then, we had three or four professionals, who first explained the pain points they had and then together we tried to define those use cases. When we defined all use cases, we prioritized them accordingly. From the point of view of the customer, they did this according to professional importance and from our perspective we also did a technical assessment, if that is feasible in the short time of the project and then together we came up with the two use cases"* (3V1). Another IBM consultant provides a similar description but identifies data as the key driving element: *"What we're saying today is: 'Start somewhere where you have a lot of data to analyze and where it's easy to put something up.' (…) That's why I'm a fan of those design thinking workshops where you have someone for two days, to jointly explore the cases, you want to approach with Watson. And then you consider what data is there, what's an employee's workday like, and what does he have to do? We identify, where can we help them (…) Then you start and see if it works and people see if it works and then you scale"* (14V1). Whereas design thinking suggests strong focus on problems and pains of individuals or groups of humans, the voices presented above show that Watson projects move the focus towards issues of technical feasibility and data.

Client-side interviewees confirm that these shifts occur – a project manager summarizes: *"I believe what comes here as a special feature is the exploratory character. (…) On the one hand you have to be explorative and, simultaneously, look at what is feasible at all, how do my data look like, how does it actually work? Then there is a lot of limitation"* (17C1). While IBM stresses the explorative character of Watson projects and brands it as design thinking, the interviews point to technology and data as constraints that limit the creative process. Design thinking imposes ambiguities regarding technology choices or solution vision to spark creativity, but AI-based application development fulfills this paradigm only to a low extent.

## Big Data Analytics

Finally, some interviewees refer to Watson as **big-data projects or data analytics**. An IBM consultant with background in data science formulates a three-stage view of Watson projects that resembles processes

*Manuscript accepted for presentation at ICIS 2022, Copenhagen*

**10**

applied in data analytics: *"So, there really were areas where we could use [it]. So, basically, we saw it as a three-step curve. (…) The first part is using cognitive intelligence to improve your data. The second component is then using cognitive intelligence in the data itself to improve it. The third part is true cognitive computing. I think that's the value chain"* (3V1). Another IBM representative argues that the particular methods for AI-based development are neither new nor sophisticated: *"It's all about patterns, big data analytics, data mining, statistics, that's not rocket science. I am sure, you've learned about those things at the university. I was at university long time ago, 97 or 98, even then, I went to the first lecture on data mining. Now comes machine learning on top to make them automatically a little smarter. The algorithms have been around for a long time, but they were just too slow"* (18V1). Some interviewees suggest one could even abstract from Watson as a specific platform and IBM as a provider. A client developer points out: *"The question is, what data do I need and what do I have to do? Some technology will be there then. Most people start with the technology first, that's wrong. It's like asking, "Is Cortana better than Siri and this better than Alexa?" That's irrelevant. The question is: What do I want to do with it and then I'll take that? Everybody's not there yet that you can use it that way. Will it ever get to that point? I don't know"* (14C1).

The interviewees who identify Watson projects as instances of data analytics emphasized what data and what analysis techniques are applied in them. However, during the project they discover that Watson projects go beyond obtaining an insight from the data which requires additional capabilities. A project manager from an insurance company outlines the necessary skill profiles: *"Of course, (…) you have to evangelize, preach what is big data analytics, what can it do and what not. It also has to be explained that this is not just about structured information, but also about unstructured information and, especially, about this keyword, cognitive computing. Underneath, people cannot imagine so much, and you have to explain that with examples (…) You need an extended team when a project arrives. You need system engineer, package infrastructure specialist, UX developer, project management. But I say, everyone has to understand machine learning (…) It's essential that they have the mindset for the cognitive aspect"* (12C1).

An IBM project manager acknowledges that data-related skills are necessary and suggests defining the role of data consultant as somebody who works with the content of the data and knows the business of the industry: *"If you then look at the project content, they [data scientists] are essential. This role of the data scientist is not needed in a normal project. I'd rather need a business analyst. (…) In a cognitive computing project, a data scientist really needs to deal with the data on the content level. You think that's something so abstract, but that's sometimes very concrete, like this example with the location of information. He has to go to the business and hear what information they need and how find that in the data. It's a very different work with data than in another projects"* (17V1). Client-side interviewees confirm that a person who combines the business-oriented and data-oriented perspective is necessary – a project manager claims: *"I am still of the opinion that I am missing a specific consulting role in this project. Someone who can talk to the business and also has a grasp of data science, data mining, cognitive technologies, and machine learning, and natural language processing and information retrieval, but with a strength in business communication"* (17C1). Overall, the archetype of data analytics puts much focus on the role of mining insight from the data and might lead to the underestimation of the necessary contacts with the business users.

### Towards a New Archetype

The data presented above shows the archetypes people initially used to inform their actions when participating in Watson projects. The statements indicate what values and objectives were driving them (e.g., developing a product vs. gaining insight) and how they impacted the way projects were organized (e.g., approaches to involve business users vs. data scientists in the project). Yet, as we noted, they found that each of those archetypes had flaws and failed to describe the project experience to the full extent. Based on the indicated shortcomings of the other paradigms, a new archetype is needed that covers the development of AI-based applications. Based on the IBM's nomenclature and the wording used by many interviewees, we refer to this new archetype as *cognitive computing*. In this section, we describe this new archetype as it emerged in the comments and opinions of the informants. Many of them express what would be needed to effectively conduct AI-application projects by indicating what needs to be changed, adapted, or added to project archetypes they knew before rather than providing a fully new recipe.

During the projects, the participants discovered that the main benefit they obtained from the projects was not an application or a data-based business insight but rather learning and researching new opportunities given the available data and the technology. A client representative formulates this as follows: *"It turned out, it was a learning process. We originally thought that the solution would be more established. But that*

*Project Archetypes for AI Development**will probably come back to their sales again. (...) But then we had to realize that a lot of questions are still unclear, even with the IBM people. Because it's like the first time that they do something in exactly that way. And that's why it was a learning process. We really had to realize that we are here in an environment where everything is shaky. (...) It's good too, that was an important experience, but that was new"* (2C1). This moves the focus of the project away from developing a solution to research. An IBM-side project manager frames it as follows: *"It's an iterative, collaborative figuring out 'what do you want, do you have the resilience, do you want to make a big bang, or do you want to do the small steps first?' Now, I have my opinion and I would say, 'Come on we'll do the small one first to get a feel for the data.' Basically, it's not that we say it scares us, we're not going to do that. It's a shared search for an optimum"* (17V1).

Accordingly, after having experienced the projects, many participants suggest starting with understanding the data first before exploring the technology or a business problem. An IBM technology consultant puts it as follows: *"'Start somewhere where you have a lot of data to analyze and where it's easy to put something up.' And then you consider what data is there, what's an employee's workday like, and what does he have to do?"* (14V1). This suggestion emerges from the experience of many projects that they underestimated the role of data quality in the initial phases of the project which prevented them from achieving the ambitious goals they established while embracing the other archetypes.

The role of data is confirmed again when the participants talk about the central artefacts relevant for the project. However, they also refer to benefit cases, Watson and Watson's abilities, and individual models. A client-side project manager explains how the recognition of technology limitations changed the way the project was conducted: *"The reason we don't have it in use is because we couldn't solve those cases I mentioned earlier with Watson. Not in the way we expected. That's why we also stopped the pilot and made a new setup. And the reason was, let's say, a big reason behind it was that we didn't structure the database the way Watson needed to, so we didn't have any databases, we didn't have any metadata and so on"* (2C1). This implies not only model training but also many steps necessary to make the training possible at all. The same manager explains how expectations mismatch the abilities of Watson: *"But that's just not the way it is. The whole data basis, the tidying up, the processes, all this is not solved by cognitive intelligence, but it starts where it actually comes along cleanly and structured and has a clean basis. And that was another problem of the project, that you have to somehow do it differently from the ground up before you can talk about cognitive intelligence"* (2C1). Given the data, data-related activities become crucial.

Yet, as indicated earlier, a data scientist who might resolve the problems around the data is not enough for the success of a Watson project. Multiple client-side project managers wished for a new role of data consultant – someone who can turn requests from the business users into adequate, data-related specifications or assess the value of existing data for solving the business users' problems. This role would supplement other roles that occur in the previous archetypes like developers, product owners, or key users.

The above statements and the overall analysis of the projects lead to further insights concerning the core assumptions about the distribution of issues and knowledge that one would have to make before entering into a Watson project. On the one hand, there is the client who has a business problem which can be solved through data-driven, probabilistic application. The client knows their business and knows most of the end-user needs, but lacks understanding of the data (despite owning the data). The vendor on the other hand provides expertise about the technology and knows about potential risks related to the data. Since both parties cannot easily assess the value of the data for AI-based development, they need to engage in a project to identify and improve applicability of the data for solving the business problem and develop an application which uses insights based on the data. A client-side manager offers the insight that engaging in Watson-based projects is a journey: "We now know we need to approach the topic step by step, we need to build know-how, we need to get the partner started and it's not a normal IT project with a release point – it's a journey. (...) That's one of the key learnings from this project. I've said that earlier, it's not just a big bang and then it's going, it's really a way. And, I think, so you can also build trust by simple cases, small cases that have only a minimal benefit, but that are already a step in this direction" (2C1). The proposed formulation of the emerging archetype for cognitive computing tries to capture this dynamic, unpredictable nature of the projects based on the reflections of the project participants.

## Discussion

Development of AI-based applications forms a new type of projects with specific requirements on how to structure, manage and evaluate them. We claim, there will be more and more of such projects given the

*Manuscript accepted for presentation at ICIS 2022, Copenhagen*
**12**

growing amounts of data and the complexity of tasks outsourced to machines. Statements collected throughout this study reveal that the project members are strongly affected by the new configuration of uncertainties as well and try to make sense of this situation while establishing patterns based on project archetypes and relating them to known processes, models, or sets of practices.

### *Existing and Emerging Archetypes*

The study participants reflect on why the initial archetypes were not fully applicable to the Watson projects. Those statements allow for the identification of an emerging archetype. Archetypes have descriptive character (Greenwood and Hinings 1993; Kirkpatrick and Ackroyd 2003; Schilling et al. 2017). Accordingly, our formulation of this archetype summarizes the observations repeated by the informants.

We agree that specific characteristics of AI such as its dependence on data, probabilistic character, or trial-and-error development strategy require a new approach (Minsky 1991; Mitchell 2019). We claim that the proposed archetype might be a starting point for creating empirically validated guidance. The advantage of using field data and creating a guidance inductively is clear: it embraces the meanings and perspectives people intuitively assigned to the events. This means, a guidance based on the proposed archetype might feel more natural to many individuals compared to guidance composed based on theoretical assumptions. We call for further research to strengthen the findings of the current study and enable a development of plausible guidance for the development of AI applications.

We need to acknowledge the positive role of archetypes initially used to make sense of the Watson projects. Many of the project members had to change their initial assumptions only late in the project and the archetypes they followed were successful at guiding their behavior. Without the archetypes it would be impossible for them to engage in any action. And given that there were just four dominating archetypes, many participants were aligned. Additionally, without the variety of approaches taken initially, the learning of the 'better' way to deal with Watson projects would have been difficult. A careful consideration of the new archetype, cognitive computing project, makes clear that it borrows elements from previous archetypes and reconfigures them. Of course, it is also unique concerning specific features like the aim or the role of data consultant, but those new values were specified against the background of existing archetypes. In other words, approaching new situations with existing artefacts is not only a natural reflex but is also helpful to identify differences and specify them accordingly. This is the blessing of project archetypes.

However, the project archetypes were radically different from each other and relied on significantly different assumptions, e.g., regarding the ultimate objective of the project is. Various individuals followed different archetypes. This is likely to generate disadvantages and requires enhanced sensemaking to understand the actions of others (Vlaar et al. 2008; Weick et al. 2005). Without an adequate frame it might be difficult to understand why some project members focus on understanding user-needs while others focus on identifying potentials of data. This calls for an explicit, more transparent, and frequent communication about the project archetypes one follows and learnings one collects during the process. This, in turn, requires sensemaking effort and resources. The stronger the previous archetypes, the more difficult it might be to change them. Accordingly, project archetype might be a curse to the project.

This twofold role of project archetypes requires special attention in research and practice. The assessment of their effectiveness for project management needs to be extended. The collected data provides a first, yet limited insight. The intuition we obtain from the interviews is that archetypes were necessary given the absence of a more adequate model. Yet, if too persistent, they can prohibit effective collaboration. Proactive communication about the archetypes might be necessary in projects relying on new technological paradigms. We claim that the vendor can take a leading role in this regard. Through actively exchanging with the clients about the learnings for project management, it can identify best practices and transfer them to further projects. This might be of more value to the clients than offering guidance on established methodologies and thus reinforcing misleading assumptions. Offering consultancy on design thinking and agile methods, IBM did enhance the confusion on the clients' side rather than highlighting IBM's competency.

This study points to the importance of project archetypes. It introduces this new concept to capture holistic understanding of project's structural and interpretive aspects. This concept extends the notion of archetype from organizational studies (Greenwood and Hinings 1988; Schilling et al. 2017). The study shows that project archetypes played a major role in establishing initial expectations towards the Watson IBM projects. The sensemaking literature left the origins of initial meaning unspecified, identifying a set of potential sources (Weick 1988; Weick et al. 2005). By focusing on projects, this study indicates the major role of project archetypes in forming the project participant's initial meanings. In other category of complex situations, individuals might tend to seek initial meanings in other type of knowledge (e.g., personal experience) or archetypes. If the defaults are known, sources of meaning might be dealt with and disconfirmed if necessary. Origins of initial meaning might become a major research area for sensemaking research.

## *Cognitive Computing as a Project Archetype*

According to the results, the practitioners did not have a clear and shared picture of what Watson-based development would involve. While some of them compared the AI platform to a word processor, others see a question-answering machine, or a digital agent to take over parts of human job, and yet others stress the fact that AI is a part of a larger infrastructure. What is most obvious just by looking at the archetypes is the diversity: whereas design thinking comes from product development (Dolata and Schwabe 2016; Uebernickel et al. 2015), customization emerged in the context of enterprise software (Hufgard and Krüger 2011), agile models were developed for overcoming difficulties in software development (Häger et al. 2015; Salo and Abrahamsson 2008), and big-data analytics is about generating insight from large amounts of structured and unstructured data (Gudivada 2017). The respondents interpret what happens in the Watson projects by choosing an archetype as a reference model: they stress the similarities, make sense of intermediate events, interpret the differences and, finally, try to transfer evaluation criteria. It is natural for humans to use analogy to make sense of novel situations that they encounter (Jensen et al. 2009; Stigliani and Ravasi 2012; Weick 1988; Weick and Sutcliffe 2015). However, since the analogous reasoning goes in many different directions for the Watson projects, we conclude that the interviewees encountered a novel phenomenon that created the need for intense sensemaking.

Some informants use *agile development* as their reference frame. On the one hand, they refer to project's high velocity and orientation at solving a problem combined with ambiguities regarding technology choices and development effort (Cockburn 2002; Larman and Basili 2003). On the other hand, they mention formal elements of the popular agile development methods like SCRUM (Schwaber and Beedle 2002). Respondents who frame the Watson-based development projects as *deployment* activities follow a simplistic view. They see creation of an AI-based application as a process of interfacing the existing data sources with the analytical machinery. Their primary concern is the choice of the right machinery. While system development models capture technical and effort-related complexities, *design thinking* was intended to manage the uncertain expectations from the stakeholders during product development (Dolata and Schwabe 2016; Häger et al. 2015; Uebernickel et al. 2015). The interviewees who follow this archetype refer to the extended emphasis on problem definition. Additionally, they confirm that the use of specific design tools breaks the complexity into manageable chunks. Finally, respondents who interpret Watson projects in the context of *big data* stress the messy data as source of similarity. However, big data and data analytics process models allot multiple steps to data understanding and composition of a consistent corpus (Azevedo and Santos 2008; Gao et al. 2015), which–as it seems–would be also necessary for IBM Watson, even though the platform offer sophisticated functionality to deal with unstructured sets. Yet, none of those frames alone can cope with the complexity that characterizes AI-based applications and their development.

Consequently, apart from listing similarities, informants point out various aspects which differentiate AI-applications development from the respective reference frames. They mention, among others, the following aspects: (1) novel performance measures and the fact that the performance varies in a manner which is hard to explain or predict (e.g., adding new data might make the model worse), (2) the limitations of the divide-and-conquer tactics which lies at the foundation of most previously known process models, (3) the need for multidimensional exploration under consideration of interdependencies between the data, preprocessing module, training or classification algorithm, and application scenario. The problem is a meta-mess: a misunderstanding of the cognitive platforms' abilities and the fit between them and the collected data. The latter leads to a messy, incomprehensible process of trial-and-error which does not only affect sub-parts of the envisioned application but lies at its core. In other words, some informants conclude that the central output of the project depends on a seemingly random and unpredictable search process.

Nevertheless, based on their experience from the analyzed Watson projects, informants establish a new archetype to capture the essence of what they encountered. We combined those statements to present a new, emerging archetype for cognitive computing. The emerging archetype is a mix between what informants observed was working in the projects and what they expect would make the projects more successful. This archetype positions the collaborative development of a probabilistic business tool between an AI platform provider and a client company at the core of the project. The progress emerges through exploration of potentials for solving an actual business problem through combination of the available data and the abilities offered by the development platform. The informants wish a data consultant to be member of those projects, i.e., a specialist who bridges the business knowledge and competency in data processing. Also, they assume that the best outcomes would be possible if the vendor had a thorough knowledge of the platform and the client of their data to engage in shared exploration of the possibilities.

We argue, the *cognitive computing* archetype, as opposite to other assumed archetypes, would have helped the informants make sense of what goes on in the projects and why. It indicates that the extensive, trial-and-error exploration is needed for finding a suitable combination between the data and features offered by the platform. It clarifies that many activities pertain to the data and that the development of the application requires collaborative efforts including collaborative engagement in data improvement. Finally, it points to the needed resources, including data, platform, benefit case, as well as skills embodied in the vision of a data consultant. Overall, the cognitive computing archetype could have prevented misunderstandings within the teams by offering a common and a more realistic reference frame.

However, the cognitive computing archetype as pictured by the informants, has some drawbacks. First, the exploratory character of the projects will still generate increased need for sensemaking and assessment: it does not provide an answer as to when the project should move from exploration to exploitation, which, in this case is preparation of the application for use by the business users. We call for more research on relation between exploration and exploitation in AI development and how practitioners should deal with the fact that there is always a possibility that another configuration of the data and the platform features might produce better results. IS is well suited to approach this challenge thanks to its experience in modelling complex systems under consideration of economic aspects (Allen and Varga 2006).

Second, the need for a data consultant indicates that there is a major knowledge deficit among business users of how AI works and what data is needed, as well as among data scientists about the processes involved in data generation and application. A data consultant seems a good kludge, but we claim that a long-term solution would involve training business professionals about basics of AI and the data scientists about challenges related to practical applications. We call for an effective training of IT specialists who can fill this gap. IS community needs to take its educational mission seriously and propagate knowledge of AI risks and opportunities to its students and the broader society (Dolata et al. 2022; Schenk and Dolata 2020).

Finally, the notion of collaboration in the proposed archetype remains underspecified. Effective collaboration depends on more than a shared understanding (Briggs et al. 2006). The cognitive computing archetype needs to be extended by collaboration techniques, processes, and tools. Design thinking as well as various agile methods offer such toolsets. Given the background in collaboration engineering (Briggs et al. 2006), IS is well equipped to propose effective collaboration patterns for the development of AI applications.

## Conclusion

AI-based application development is on the increase. Yet, no methodology has reached the status of a dominating approach. Lacking a clear guidance, project members act based on the available project archetypes. This article presents those archetypes indicating what aspects of the AI development platforms or the projects invoke those archetypes. It makes the following contributions: First, it introduces and defines the notion of *project archetypes* which can be used for analysis of projects in various contexts. Based on the data analysis, it shows which dimensions are useful to describe the project archetypes (aims, artefacts, activities, abilities, assumptions). It also provides theoretical underpinning for the concept of project archetypes by rooting it in the sensemaking literature. Second, the paper shows that a new archetype for *cognitive computing* is emerging in a bottom-up manner, lists its limitations, and calls for IS research concerning exploration, collaboration, and education to improve the success of AI development projects. In this paper, researchers obtain a first insight into the issues related to the management of AI-based projects. Practitioners can benefit from the learnings other projects have already made and become aware of potential difficulties. They also learn to actively approach project archetypes in project teams rather than assuming alignment.

Those insights come with limitations. The cases we studied were all between companies located in Switzerland and the Swiss IBM branch. The results need to be replicated for external validity. The overall number of cases is limited too. The ambition to interview a company representative and an IBM representative for each case limited our choice, because some projects were highly confidential and company representative feared leaks. Also, interviews with several project members from each side could provide even more discursive material and provide essential hints on the management, e.g., on the communication of goals and context. We hope, to approach those limitations in further research.

## Acknowledgement

We express our best gratitude to Daniel Oettli and Nicola Storz for their engagement in collecting the data which form the basis for this manuscript. We also thank Dr. Alain Gut and Philip Spaeti for their fundamental support throughout the research project.

*Project Archetypes for AI Development*Kim, M., Zimmermann, T., DeLine, R., and Begel, A. 2016. "The Emerging Role of Data Scientists on Software Development Teams," in *Proc. Intl. Conf. Software Engineering*, ACM, pp. 96–107.
Kirkpatrick, I., and Ackroyd, S. 2003. "Archetype Theory and the Changing Professional Organization: A Critique and Alternative," *Organization* (10:4), SAGE Publications Ltd, pp. 731–750.
Konar, A. 1999. *Artificial Intelligence and Soft Computing: Behavioral and Cognitive Modeling of the Human Brain*, CRC Press.
Larman, C., and Basili, V. R. 2003. "Iterative and Incremental Developments," *Computer* (36:6).
Larson, E. J. 2021. *The Myth of Artificial Intelligence: Why Computers Can't Think the Way We Do*, Cambridge, Massachusetts: The Belknap Press of Harvard University Press.
Laughlin, R. C. 1991. "Environmental Disturbances and Organizational Transitions and Transformations: Some Alternative Models," *Organization Studies* (12:2), pp. 209–232.
Madhavji, N. H., Miranskyy, A., and Kontogiannis, K. 2015. "Big Picture of Big Data Software Engineering: With Example Research Challenges," in *Proc. Intl. WS BIG Data Software Engineering*, IEEE Press.
McAfee, A., and Brynjolfsson, E. 2017. *Machine, Platform, Crowd: Harnessing Our Digital Future*, New York: W.W. Norton & Company.
Minsky, M. 1991. "Logical vs. Analogical or Symbolic vs. Connectionist or Neat vs. Scruffy," in *Artificial Intelligence at MIT Expanding Frontiers*, Cambridge, MA, USA: MIT Press, pp. 218–243.
Mitchell, M. 2019. *Artificial Intelligence: A Guide for Thinking Humans*.
Möller, F., Bauhaus, H., Hoffmann, C., Niess, C., Otto, B., and Isst, F. 2019. "Archetypes of Digital Business Models in Logistics Start-UPS.," in *Proc. European Conf. Information Systems*.
Molnar, C. 2020. *Interpretable Machine Learning*, Leanpub.
Muller, M., Lange, I., Wang, D., Piorkowski, D., Tsay, J., Liao, Q. V., Dugan, C., and Erickson, T. 2019. "How Data Science Workers Work with Data" in *Proc. Conf. Human Factors in Computing Systems*, ACM.
Otero, C. E., and Peter, A. 2015. "Research Directions for Engineering Big Data Analytics Software," *IEEE Intelligent Systems* (30:1), pp. 13–19.
Pearl, J., and Mackenzie, D. 2018. *The Book of Why: The New Science of Cause and Effect*, NY: Basic Books.
PMI (ed.). 2017. *A Guide to the Project Management Body of Knowledge / Project Management Institute*, (Sixth edition.), PMBOK Guide, Newtown Square, PA: Project Management Institute.
Pratt, M. G. 2000. "The Good, the Bad, and the Ambivalent: Managing Identification among Amway Distributors," *Administrative Science Quarterly* (45:3), SAGE Publications Inc, pp. 456–493.
Reamy, T. 2016. *Deep Text: Using Text Analytics to Conquer Information Overload, Get Real Value from Social Media, and Add Big(Ger) Text to Big Data*, Medford, New Jersey: Information Today, Inc.
Runeson, P., and Höst, M. 2009. "Guidelines for Conducting and Reporting Case Study Research in Software Engineering," *Empirical Software Engineering* (14:2), p. 131.
Salo, O., and Abrahamsson, P. 2008. "Agile Methods in European Embedded Software Development Organisations: A Survey" *IET Software* (2:1), pp. 58–64.
Sawyer, S. 2004. "Software Development Teams," *Communications of the ACM* (47:12), pp. 95–99..
Schenk, B., and Dolata, M. 2020. "Facilitating Digital Transformation through Education: A Case Study in the Public Administration," in *Proc. Hawaii Intl. Conf. System Sciences*.
Schilling, R. D. 2018. "Theories to Understand the Dynamic Nature of Enterprise Architecture," in *IEEE International Enterprise Distributed Object Computing Workshop (EDOCW)*, IEEE, pp. 153–161.
Schilling, R. D., Haki, M. K., and Aier, S. 2017. "Introducing Archetype Theory to Information Systems Research: A Literature Review and Call for Future Research," in *Proc. Conf. Wirtschaftsinformatik*.
Schwaber, K., and Beedle, M. 2002. *Agile Software Development with Scrum*, (Vol. 1), Prentice Hall.
Stigliani, I., and Ravasi, D. 2012. "Organizing Thoughts and Connecting Brains: Material Practices and the Transition from Individual to Group-Level Prospective Sensemaking," *Acad. of Managment J*. (55:5).
Uebernickel, F., Brenner, W., Naef, T., Pukall, B., and Schindlholzer, B. 2015. *Design Thinking: Das Handbuch*, Frankfurt am Main: Frankfurter Allgemeine Buch.
Vlaar, P. W. L., van Fenema, P. C., and Tiwari, V. 2008. "Cocreating Understanding and Value in Distributed Work" *MIS Quarterly* (32:2), p. 227.
Weick, K. E. 1988. "Enacted Sensemaking in Crisis Situations," *Journal of Management Studies* (25:4).
Weick, K. E. 1995. *Sensemaking in Organizations*, Foundations for Organizational Science, SAGE.
Weick, K. E., and Sutcliffe, K. M. 2015. *Managing the Unexpected: Sustained Performance in a Complex World*, (Third edition.), Hoboken, NJ: John Wiley & Sons, Inc.
Weick, K. E., Sutcliffe, K. M., and Obstfeld, D. 2005. "Organizing and the Process of Sensemaking," *Organization Science* (16:4), INFORMS, pp. 409–421.
Winter, R. 2016. "Establishing 'Architectural Thinking'in Organizations," in *Proc. IFIP Working Conf. Practice of Enterprise Modeling*, Springer, pp. 3–8.
Yin, R. K. 2003. "Applications of Case Study Research," *Series, 4th. Thousand Oaks: Sage Publications*.*Manuscript accepted for presentation at ICIS 2022, Copenhagen*
**17**